\documentclass[aps,prl,twocolumn,floatfix,letterpaper,superscriptaddress,amssymb,eprint,longbibliography]{revtex4-2}
\usepackage[latin9,utf8]{inputenc}
\usepackage{stmaryrd}
\usepackage{graphicx}
\usepackage{multirow}
\usepackage{mathtools}
\usepackage{xcolor}
\usepackage{textcomp}
\usepackage{gensymb}
\usepackage{calligra}
\usepackage{braket}
\usepackage{float}
\usepackage{bbold}
\usepackage{verbatim}
\usepackage{soul} 

\usepackage{xcolor,mathrsfs,dsfont}
\definecolor{darkblue}{RGB}{0,0,150}
\definecolor{nightblue}{RGB}{0,0,100}
\definecolor{evergreen}{RGB}{0,120,0}
    
\usepackage{graphicx,mathtools,bm}
\graphicspath{{./Figures/}}
\usepackage[
colorlinks,
citecolor=blue,
linkcolor=blue,
urlcolor=nightblue]{hyperref}

\let \oldbm \bm
\renewcommand{\vec}[1]{\oldbm{#1}}

\usepackage[english]{babel}
\usepackage[babel,kerning=true,spacing=true]{microtype}
\usepackage[utf8]{inputenc}

\definecolor{DarkRed}{RGB}{100,0,0}

\def\bm{{\vec m}}

\makeatother

\begin{document}
\title{Non-Hermitian magnetic moment}
\author{Bar Alon}
\author{Moshe Goldstein}
\affiliation{Raymond and Beverly Sackler School of Physics and Astronomy, Tel Aviv University, Tel Aviv 6997801, Israel}
\author{Roni Ilan}

\affiliation{Raymond and Beverly Sackler School of Physics and Astronomy, Tel Aviv University, Tel Aviv 6997801, Israel}
\affiliation{Simons Emmy Noether Fellow, Perimeter Institute for Theoretical Physics, 31 Caroline Street North, Waterloo, Ontario, Canada N2L 2Y5}


\begin{abstract}
We construct a semiclassical theory for electrons in a non-Hermitian periodic system subject to perturbations varying slowly in space and time. We derive the energy of the wavepacket to first order in the gradients of the perturbations. Applying the theory to the specific case of a uniform external magnetic field, we obtain an expression for the orbital magnetization energy. Using the principles of non-Hermitian dynamics, we define a physically meaningful non-Hermitian generalization of the angular momentum operator and show that it is compatible with the real part of the orbital magnetic moment. The imaginary part of the orbital magnetic moment is also discussed and shown to originate from an imaginary counterpart to the angular momentum that gives rise to a non-Hermitian generalization of the Aharonov-Bohm effect.
\end{abstract}

\maketitle

\emph{Introduction.} 
Semiclassical band theory is vital in the study of electron dynamics in periodic media \cite{ashcroft1976solid}. Within this method, a narrow wave packet is constructed as a superposition of Bloch states, and its properties under the influence of classical external perturbing fields are studied. The topological and geometric properties of the Bloch bands are known to have far-reaching consequences on wavepacket dynamics and response to perturbations. In particular, these properties are essential to the modern theory of polarization \cite{VanderbiltPhysRevB.47.1651,OrtizPhysRevB.49.14202,RestaRevModPhys.66.899}, orbital magnetization \cite{chang_berry_1996,ThonhauserPhysRevLett.95.137205}, and anomalous thermoelectric transport \cite{chien1980hall,OnodaQuantum2008PhysRevB.77.165103} and are responsible for the appearance of the anomalous velocity \cite{chang_berry_1996,sundaram_wave-packet_1999,chang_berry_2008,xiao_berry_2010} giving rise to the anomalous Hall effect \cite{PhysRevLett.88.207208,NagaosaAnomalousRevModPhys.82.1539}.
When the external perturbation is in the form of a magnetic field, the semiclassical theory predicts additional effects, such as magnetotransport \cite{XiaoDOSPhysRevLett.95.137204,xiao_berry_2010}, valley polarization \cite{xiao_valley-contrasting_2007}, and the de Haas-van Alpen effect \cite{ashcroft1976solid}.

Recently, there has been growing interest in extending existing theories to apply to non-Hermitian Hamiltonians \cite{Bender_2007,ShenPhysRevLett.120.146402,martinezalvarezTopologicalStatesNonHermitian2018,Ghatak_2019,KunstPhysRevB.99.245116,gong_topological_2018,ashida_non-hermitian_2020, MartinezTopological2024} that may provide an effective description of non-equilibrium processes in a wide array of fields, ranging from biological \cite{nelson_non-hermitian_1998,lubensky_pulling_2000} to nuclear physics \cite{feshbach_model_1954,feshbach_unified_1958}. Such Hamiltonians have already been realized in a variety of systems, including mechanical \cite{Brandenbourger_robotic_2019,Ghatak_mechanical_2020} and electrical \cite{Helbig_Generalized_2020,Hofmann_2020_PhysRevResearch.2.023265,zou_observation_2021} metamaterials, as well as in photonics \cite{Weidemann_topological_2020,Xiao_bulk-boundary_2020,XiaoPT_2021PhysRevLett.126.230402,wang_non-hermitian_2023}, magnonics \cite{hurst_non-hermitian_2022,yu_non-hermitian_2024}, and ultracold atomic gases \cite{Liang2022PhysRevLett.129.070401}. Moreover, synthetic gauge potentials and magnetic fields have already been proposed and realized in some of these platforms \cite{Lau_coldatomAB_2023,Lin_synthMagColdatoms_2009,Dalibard_artificialgaugeAtoms_2011,Meir_temporalAB_2025}.
Consequently, there is ample motivation to develop a comprehensive theory of charged particles whose dynamics are governed by a non-Hermitian Hamiltonian and are subject to magnetic fields. 

Recent works have investigated the effects of both real \cite{shen_quantum_2018,okuma_topological_2019} and imaginary magnetic fields \cite{ozawa_two-dimensional_2024} in certain non-Hermitian models, particularly in relation to the non-Hermitian skin effect \cite{lu_magnetic_2021,li_enhancement_2023}. Additionally, methods for the realization of pseudomagnetic fields in non-Hermitian systems have been proposed \cite{zhang_non-hermitian_2020}. However, a general non-Hermitian theory of the motion of a charged particle in a magnetic field has yet to be formulated. In particular, there has been no non-Hermitian counterpart to the Hermitian semiclassical theory that provides a closed form for the magnetic moment of a Bloch electron's wavepacket and ties it to its orbital angular momentum \cite{chang_berry_1996,chang_berry_2008}. 
To fill this gap, we construct the general theory for the energy of a wavepacket propagating in a slowly perturbed periodic crystal. Applying this theory to the specific case of magnetic vector potentials that vary slowly in space, we derive the energy of the wavepacket to first order in the field. We then find an expression for the orbital magnetic moment of the wavepacket and show that it is composed of two contributions. The first couples the anomalous non-Hermitian Berry connection \cite{xu_weyl_2017} to the Lorentz force, similarly to known results for electric fields \cite{silberstein_berry_2020,alon_quantum_2024}. The second is related to the self-rotation of the wavepacket about its center. We discuss the relation between the magnetic moment and the orbital angular momentum of the wavepacket and show that the physical angular momentum operator, when properly defined in the non-Hermitian setting, is responsible for the real part of the orbital moment. We also show that the imaginary part of the moment is determined by an imaginary orbital momentum that gives rise to a non-Hermitian generalization of the Aharonov-Bohm effect \cite{Aharonov_Bohm_1959}.

\emph{Non-Hermitian Bloch bands.} 
We begin by considering the problem of a physical system that evolves in time according to the Schr\"{o}dinger equation 
\begin{equation}
     \frac{d}{dt}\ket{\psi(t)} = -iH(t)\ket{\psi(t)}
\end{equation}
for some time-dependent Hamiltonian operator $H(t)$. We assume throughout this Letter that the Hamiltonian is the sum of two parts, $H(t)=H_p(t)+\Delta H(t)$, where $H_p(t)$ is non-Hermitian and periodic in real space with respect to some (time-independent) lattice while $\Delta H(t)$ is a small nonperiodic addition. We also assume the absence of spin-orbit coupling. Under these assumptions, we may construct the time-dependent eigenbasis of $H_p(t)$ according to Bloch's theorem $\ket{\psi_{nk}(t)}$.

Since $H_p(t)$ is non-Hermitian, its energies $\varepsilon_{nk}(t)$ are, in general, complex numbers. The real part of the energy determines the rate at which stationary states accumulate phase, as in the Hermitian case. The imaginary part determines the gain or loss of the state. In a superposition, the weight of different eigenstates may be amplified or suppressed relative to each other. This creates a natural preference for \emph{maximally amplified} states, namely those with the largest imaginary energies, and leads to fundamental restrictions to the applicability of the adiabatic theorem  \cite{berry_slow_2011,ibanez_adiabaticity_2014,wang_non-hermitian_2018,silberstein_berry_2020}.

In many cases, the basis formed by the eigenfunctions of $H_p$ may not be orthogonal. Specifically, this occurs whenever $H_p$ does not commute with its Hermitian conjugate (i.e., it is not normal). In such cases, it is important to distinguish the eigenbasis of $H_p$, also known as the \emph{right} eigenbasis $\ket{\psi^R_{nk}(t)}$, from its dual basis, also known as the \emph{left} eigenbasis $\ket{\psi^L_{nk}(t)}$, which is defined by the product
\begin{equation}
     \label{eq:def_leftright}
    \braket{\psi^L_{n^\prime k^\prime} (t)| \psi^R_{nk}(t)} = \delta_{nn^\prime}\delta(k-k^\prime).
\end{equation}
In contrast, the product between two right eigenstates is given by the Gram matrix
\begin{equation}
    \label{eq:grammian}
    I_{n^\prime k^\prime;nk}(t) = \braket{\psi^R_{n^\prime k^\prime} (t)| \psi^R_{nk}(t)} = I_{n^\prime n}(k,t)\delta(k-k^\prime).
\end{equation}
Note that due to Bloch's theorem, the Gram matrix is always diagonal in $k$, but not necessarily in $n$. 

Unlike the right eigenstates, the left eigenstates are not necessarily stationary states of $H_p$. Nevertheless, they play a vital role in the dynamics, control the response of the system to perturbations and adiabatic changes \cite{ShenPhysRevLett.120.146402,xu_weyl_2017,schomerus_nonreciprocal_2020,silberstein_berry_2020,hu2024rolequantumgeometrictensor}, and are necessary for defining physical quantities of import, such as the non-Hermitian Berry phase \cite{GDattoli_1990,longhi_complex_2023,PhysRevResearch.5.L032026}. The non-orthogonality of modes and the resulting distinction between left and right states expresses itself through a distinction between two different generalizations of the Berry connection
\begin{equation}
    \label{eq:Berry_connections}
    A^{RR}_\lambda = i\frac{\braket{u^R_{nk}(t)|\partial_\lambda u^R_{nk}(t)}}{I_{nn}(k,t)},\quad A^{LR}_\lambda = i\braket{u^L_{nk}(t)|\partial_\lambda u^R_{nk}(t)},
\end{equation}
with $\lambda$ an arbitrary system parameter. We note that the first type of Berry connection, $A^{RR}$, involves only the eigenstates of the $n$th band and their derivatives. Moreover, its integrals over closed paths in parameter space are real, and its curl yields a Berry curvature which appears in the anomalous velocity \cite{xu_weyl_2017,silberstein_berry_2020,alon_quantum_2024}. This connection preserves many of the basic properties of the Berry connection of Hermitian systems. The second type of Berry connection, $A^{LR}$, involves both left and right states and always depends on all bands. This connection quantifies the response of the system to external perturbations, and its integral over closed paths in parameter space yields the non-Hermitian counterpart of the Berry phase \cite{GDattoli_1990,longhi_complex_2023,PhysRevResearch.5.L032026}. This connection, as well as the Berry phase it induces, is generally complex. Remarkably, the difference between the connections $A^{LR}-A^{RR}$ is a gauge invariant quantity \cite{xu_weyl_2017,silberstein_berry_2020,PhysRevResearch.5.L032026}. Moreover, it plays an important role in non-Hermitian perturbation theory and the resulting dynamics \cite{silberstein_berry_2020,alon_quantum_2024,hu2024rolequantumgeometrictensor}, as we also show in the following sections.
Finally, we note the existence of systems with exceptional points \cite{Heiss_2012,
AliMiriExceptional,bergholtz_exceptional_2021}, which occur whenever the Hamiltonian fails to be diagonalizable. We do not consider such systems in this Letter.

\emph{Time-dependent single-band wave-packets.} 
Consider a single-band wavepacket constructed from Bloch eigenstates of the periodic Hamiltonian
\begin{equation}
    \ket{W}=\int_k w_{nk}\ket{\psi^R_{nk}}.
\end{equation}
We define the envelope function to be
\begin{equation}
    \label{eq:def_envelope}
    g(k)=\vert w_{nk}\vert^2I_{nn}(k).
\end{equation}
We require that the wavepacket is well localized in $k$ space with a well-defined central momentum $k_c$, and take the limit of a sharply peaked envelope function
\begin{equation}
\label{eq:peakCondition}
    g(k) \sim N\delta(k-k_c),
\end{equation}
where $N=\braket{W|W}$ is the squared norm of the wave packet state. Note that all quantities are explicitly time dependent, although we have suppressed this in the notation.

If the packet is confined to a maximally amplified band, the non-Hermitian adiabatic theorem may be applied \cite{GNenciu_1992,silberstein_berry_2020}, provided the rate of change of $H_p(t)$ and any external perturbations are sufficiently small. Under these conditions, the dynamics of the packet are well approximated by the Schr\"{o}dinger equation projected onto the dominant band:
\begin{equation}
    \label{eq:SingleBand_Schrodinger}
    \dot{w}_{nk} = -\int_{k'}w_{nk'} \left( i\bra{\psi^L_{nk}}H \ket{\psi^R_{nk'}}+\braket{\psi^L_{nk}|\partial_t\psi^R_{nk'}}\right),
\end{equation}
where we have taken the explicit time dependence of the eigenstates into account.

To obtain the effective energy of the wavepacket, we calculate the expectation value of the full Hamiltonian as
\begin{equation}
    \label{eq:Energy_eff_def}
    \mathcal{E} = \frac{\bra{W}H\ket{W}}{\braket{W|W}}.
\end{equation}
By writing the Hamiltonian as $H=i(d/dt)$ and making use of the band-projected Schr\"{o}dinger equation \eqref{eq:SingleBand_Schrodinger}, we may derive a simple expression for the single-band effective energy (see Supplemental Material):
\begin{equation}
    \label{eq:Energy_eff}
    \mathcal{E} = \langle P_n H \rangle_W - (A^{LR}_t - A^{RR}_t).
\end{equation}
The first term is then the expectation value of the band projected (full) Hamiltonian $P_nH$ with respect to the wavepacket state $\ket{W}$, where we have defined the non-Hermitian band projection operator
\begin{equation}
    \label{eq:proj_def}
    P_n = \int_k \ket{\psi^R_{nk}}\bra{\psi^L_{nk}}.
\end{equation}
We note that this is indeed a projection operator which annihilates all but the $n$th band of the periodic Hamiltonian $H_p(t)$. The second term is precisely the difference between the Berry connections with time as their adiabatic parameter. A special case of note is when the left and right eigenstates of the $n$th band coincide. In this case the projector $P_n$ becomes Hermitian and acts trivially on the state to its left. Furthermore, the difference between Berry connections vanishes, and the effective energy becomes simply the expectation value of the full Hamiltonian. This occurs precisely when the $n$th band is orthogonal to all others.

We remark that Eq. \eqref{eq:Energy_eff} may be used to rederive the effective energy in the presence of an electric field, realized through a time-dependent vector potential~\cite{alon_quantum_2024}.

\emph{Quasi-local Hamiltonians.} 
Following the approach of Refs. \cite{sundaram_wave-packet_1999,niu_electron_2001}, we consider a periodic Hamiltonian which is perturbed by a modulating spatially dependent external field $F(\hat{r})$. Since the Hamiltonian is periodic except for its dependence on the modulating field, translational symmetry is recovered if the field is constant in space. Explicitly, if $F(\hat{r})=c$, then $H(F(\hat{r}))=H(c)$ retains the periodicity of the lattice, and we may write a corresponding Bloch Hamiltonian $H(k,c)$. Going beyond this trivial case, for fields that change slowly with respect to the spread of a wavepacket, one may approximate the field as a constant by taking it to be $F(\hat{r})=F(r_{c})$, where $r_{c}$ is the center-of-mass position of the wavepacket. This results in the approximation $H(\hat{r})\approx H(r_{c})$. The variation of the field across space can then be taken into account up to first order by linearizing the Hamiltonian around the central position $r_{c}$ \cite{sundaram_wave-packet_1999},
\begin{equation}
    \label{eq:quasilocal_Ham}
    H(\hat{r})\approx H(r_{c})+\frac{1}{2}\left[(\hat{r}-r_{c})\frac{\partial H}{\partial r_{c}}+\frac{\partial H}{\partial r_{c}}(\hat{r}-r_{c})\right],
\end{equation}
where the derivatives are also taken at the wavepacket center $r_{c}$. This is the quasilocal Hamiltonian.

The above Hamiltonian differs from the unperturbed system not only by the second term of Eq. \eqref{eq:quasilocal_Ham}, but also acquires an implicit time dependence due to the dependence on $r_c$, which changes as the wavepacket propagates through the system. Both of the above perturbative effects scale with the spatial derivatives of the modulating field. When considering magnetic fields, the modulating field in the Hamiltonian is the magnetic vector potential $\mathcal{A}(\hat{r})$, and thus the perturbation is controlled by the strength of the magnetic field $B=\nabla\times\mathcal{A}$.

Due to the implicit time dependence of the quasilocal Hamiltonian, we use Eq. \eqref{eq:Energy_eff} to derive a general expression for the effective energy of the wavepacket. This derivation yields (see Supplemental Material)
\begin{equation}
    \label{eq:Energy_quasilocal}
    \begin{split}
        \mathcal{E}&=\varepsilon_{nk}(r_{c})-\dot{r}_c\cdot(A_{r_c}^{LR}-A_{r_c}^{RR})+\nabla_{r_c}\varepsilon_{nk}\cdot(A_{k}^{LR}-A_{k}^{RR})\\
        &+\frac{i}{2}\left[\bra{\nabla_{r_{c}}u_{nk}^{L}}\cdot(\varepsilon_{nk}-H(k))\ket{\nabla_{k}u_{nk}^{R}}\right.\\
        &\left.-\bra{\nabla_{k}u_{nk}^{L}}\cdot(\varepsilon_{nk}-H(k))\ket{\nabla_{r_c}u_{nk}^{R}}\right],
    \end{split}
\end{equation}
where all values are taken at the central momentum of the wavepacket $k_c$. Following the unperturbed band energy at the wavepacket center are two dual terms. The first of these terms couples the velocity to the difference between Berry connections with position as their parameter. This term follows from the difference between Berry connections in Eq. \eqref{eq:Energy_eff} and the implicit time dependence of $r_c$. The second of the dual terms couples the gradient of the energy, which we think of as a generalized complex-valued force, with the difference between Berry connections with crystal-momentum as their parameter. This term is part of the contribution which originates from the second term in \eqref{eq:quasilocal_Ham}. Both of the above terms vanish in the Hermitian case where the left and right states coincide and are therefore unique to the non-Hermitian setting. The last two terms involve derivatives of the eigenstates with respect to both the crystal momentum and the central position, and it also originates from the gradient term in the quasilocal Hamiltonian. Unlike the previous terms, this term does not vanish in the Hermitian case. As we show in the next section, this term is responsible for the magnetic moment of the Hermitian setting.

\emph{Magnetic moment.} 
We now employ the quasi-local Hamiltonian formalism of the previous section to the specific case of constant magnetic fields. We work in the symmetric gauge. By the Peierls substitution the local periodic Hamiltonian is then given by
\begin{equation}
    \label{eq:Hamloc_mag}
    H(r_c)=e^{ikr}H_0(k+\frac{1}{2}B\times r_c)e^{-ikr},
\end{equation}
where $H_0(k)$ is the unperturbed Bloch Hamiltonian. This problem is most easily treated in terms of the gauge-invariant momentum, $p(k,r_c)=k+\frac{1}{2}B\times r_c$, where the Bloch eigenfunctions of the local Hamiltonian attain a simple form in terms of the eigenfunctions of the Hamiltonian in the absence of fields.
\begin{equation}
    \ket{u^R_{nk}(r_c)}=\ket{u^R_{np}},\quad \ket{u^L_{nk}(r_c)}=\ket{u^L_{np}}.
\end{equation}
One may then write the full Hamiltonian in the quasilocal approximation in the Bloch representation,
\begin{equation}
    \label{eq:HamQuasiloc_mag}
    H(k) = H_0(p) 
    +B\cdot(\frac{1}{2}(\hat{r}-r_{c})\times\nabla_p H_0).
\end{equation}
To make use of the results of the previous section, we must evaluate the derivatives with respect to $r_c$ and $k$. We do this by representing both these derivatives in terms of the gauge-invariant momentum $p$ as
\begin{equation}
    \nabla_{r_c}\to -\frac{1}{2}B\times\nabla_p,\quad \nabla_k\to\nabla_p.
\end{equation}
Inserting the above into the effective energy \eqref{eq:Energy_quasilocal} yields the expression for the effective energy in a magnetic field
\begin{equation}
    \label{eq:Energy_mag}
    \begin{split}
        \mathcal{E}_M &= \varepsilon_{np} 
        +\frac{1}{2} (\dot{r}\times B)\cdot (A^{LR}_p-A^{RR}_p) \\
        &-  B\cdot
        \left[\frac{1}{2}\nabla_p\varepsilon_{np}\times (A^{LR}_p-A^{RR}_p) \right.\\ 
        &+ \left.\frac{i}{2} \bra{\nabla_p u_{np}^{L}}\times(\varepsilon_{np}-H(p))\ket{\nabla_p u_{np}^{R}}
        \right].
    \end{split}
\end{equation}
This is the main result of this Letter. The energy in a magnetic field is composed of the unperturbed band energy supplemented by two additional terms that are linear in the magnetic field. The first originates from the adiabatic correction to the time-dependent single-band approximation, and is proportional to the product of the Lorentz force and the difference of Berry connections. This is reminiscent of non-Hermitian energy corrections in the presence of electric fields, which similarly couple the difference of Berry connections to the electric field \cite{silberstein_berry_2020,alon_quantum_2024}.
The second term is Zeeman-like. It originates from the quasilocal addition to the band Hamiltonian and is related to the self-rotation of the wavepacket about its center of mass. In the Hermitian case, this term simplifies to the usual orbital magnetic moment. This is not surprising given the form of the magnetic quasilocal term of the Hamiltonian \eqref{eq:HamQuasiloc_mag}, which parallels the Zeeman term $B\cdot L$ of the Hermitian case \cite{chang_berry_1996,chang_berry_2008}. In the general non-Hermitian case however, this term becomes complex valued. Its relation to the orbital angular momentum of the wavepacket is discussed in the following section.

\emph{Orbital angular momentum.} 
We now generalize the orbital angular momentum operator to the non-Hermitian setting and relate it to the magnetic moment. A naive definition for the orbital momentum operator is given by
\begin{equation}
    \label{eq:AM_nh}
    \hat{\mathcal{L}}(k) = \frac{1}{2}(\hat{r}-r_{c})\times\nabla_k H.
\end{equation}
This operator reduces to (and is identical in form to) the orbital angular momentum operator of Hermitian physics. Moreover, as is seen in Eq. \eqref{eq:HamQuasiloc_mag}, it appears from the coupling of the Hamiltonian to magnetic fields. This operator, however, is not Hermitian in general, making it unsuitable for describing a real measurable property such as the angular momentum. This occurs because the orbital angular momentum derives from the velocity opeartor. In the Hermitian case the velocity operator is defined by $\hat{v}=-i[r,H]$, which has the Bloch form
\begin{equation}
    -ie^{-ikr}[r,H]e^{ikr}=\nabla_k H.
\end{equation}
This definition for the velocity operator also does not yield a Hermitian operator in general when applied to a non-Hermitian Hamiltonian and thus cannot represent a measurable quantity. By extension, every operator constructed using $\nabla_k H$ is also expected to become unphysical in the non-Hermitian setting.

Recalling that the use of the commutator in the definition of $\hat{v}$ derives from the Ehrenfest theorem suggests that in order to derive a physically meaningful angular momentum operator one should construct it by making explicit use of the principles of non-Hermitian dynamics. The non-Hermitian generalization of the Ehrenfest theorem \footnote{Here we use the term 'Ehrenfest Theorem' to refer to the equation of motion for the expectation value of any operator, not necessarily position or momentum.} (for a time-independent operator) is given by \cite{alon_quantum_2024}
\begin{equation}
    \label{eq:NH_Ehrenfest}
    \frac{d}{dt}\langle O\rangle=-i\left[ \langle OH- H^\dagger O\rangle - \langle H-H^\dagger\rangle \langle O\rangle\right].
\end{equation}
Applying this to the position operator $\hat{r}$, the new definition for the velocity operator becomes
\begin{equation}
    \label{eq:NH_velocity}
    \hat{v}=-i\left[(\hat{r}-r_c)H- H^\dagger (\hat{r}-r_c)\right].
\end{equation}
This operator is Hermitian, as is necessary for a physical velocity, and reduces to the usual expression for the velocity operator when $H$ is Hermitian. Moreover, it yields the correct group velocity, $\text{Re}\nabla_k \varepsilon(k)$, when applied to states localized in momentum space. We therefore conclude that it is a reasonable candidate for a physical velocity operator.

The physical angular momentum is then defined as in the Hermitian case by taking the cross product of the position and the velocity 
\begin{equation}
    \label{eq:AM_def}
    \hat{L}=\frac{1}{2}[(\hat{r}-r_c)\times \hat{v} - \hat{v} \times (\hat{r}-r_c)].
\end{equation}
We have symmetrized the cross product since the generalized form of the velocity does not necessarily commute with the position. This definition gives a Hermitian operator that reduces to the usual expression for the angular momentum operator when the Hamiltonian is Hermitian, as desired.

The relation between the physical angular momentum defined above and the operator \eqref{eq:AM_nh} appearing in the magnetic moment becomes apparent when one considers their Cartesian components
\begin{equation}
    \label{eq:AM_coords}
    \hat{L}_i = i\varepsilon_{ijk}(\hat{r}-r_c)_j\frac{H+H^\dagger}{2}(\hat{r}-r_c)_k,
\end{equation}
\begin{equation}
    \label{eq:AM_nh_coords}
    \hat{\mathcal{L}}_i = i\varepsilon_{ijk}(\hat{r}-r_c)_j H(\hat{r}-r_c)_k.
\end{equation}
One may therefore conclude that the physical angular momentum $\hat{L}$ is precisely the real part of the magnetic moment, $\text{Re}\hat{\mathcal{L}} = (\hat{\mathcal{L}}+\hat{\mathcal{L}}^\dagger)/2$. Put differently, we may think of the magnetic moment appearing in the magnetic Hamiltonian \eqref{eq:HamQuasiloc_mag} as the sum of two contributions,  real and imaginary. The first is the physical angular momentum, which changes the real part of the energy linearly in the magnetic field, as is familiar from the Hermitian setting. The second is then an `imaginary angular momentum,' which similarly couples to the magnetic field and affects the imaginary part of the energy. That is, the imaginary angular momentum generates gain (or loss) instead of energy in the usual sense. From the above expressions we may also glean another remarkable property of the physical angular momentum, namely, that it depends only on the \emph{real} part of the Hamiltonian $\text{Re}H$. That is, changes to the imaginary part of $H$ do not alter the angular momentum of any given state (although its time evolution still depends implicitly on the whole of $H$ through the time evolution of the state, as with all observables). This is unusual because, while the overall dynamics of all states depend on both the Hermitian and the anti-Hermitian parts of $H$, their self-rotation, which is a type of motion, does not. As a contrasting example, the velocity operator \eqref{eq:NH_velocity} does not share this property. 

It remains to gain an understanding of the `imaginary angular momentum' and the physical effect it represents. We do this by considering the idea of a quantum state with an imaginary velocity, that is, a state which evolves in time approximately as $\bar{\psi}(x)\to\bar{\psi}(x+ivt)$, where we understand $\bar{\psi}$ to mean an analytic continuation of the wavefunction $\psi$. The Cauchy-Riemann equations for $\bar{\psi}$ are then given by
\begin{subequations}
\label{eq:CR_imagVelocity}
\begin{align}
    \partial_t \varphi&= v\cdot \nabla_x\ln{\vert{\bar{\psi}}\vert}, \\
    \partial_t \ln{\vert{\bar{\psi}}\vert} &= -v\cdot \nabla_x \varphi,
\end{align}
\end{subequations}
where $\varphi$ is the phase of the quantum state. An important consequence of the Cauchy-Riemann equations is that the norm of the state decays (grows), i.e. loss (gain) is generated, whenever the imaginary velocity is (anti-)aligned with space gradients of the phase of the quantum state. An interpretation of the 'imaginary angular momentum' and its appearance in the magnetic Hamiltonian now becomes apparent. On the one hand, the angular momentum operator measures the rotation of the velocity field about the center of the wavepacket. The 'imaginary angular momentum' therefore measures the tendency of the state to experience dissipation due to phase rotation about its center. On the other hand, in the presence of a vector potential, the spatial gradients in Eq. \eqref{eq:CR_imagVelocity} become gauge-invariant derivatives, $\nabla_x \to \nabla_x - \mathcal{A}$, in accordance with the Aharonov-Bohm effect. The `imaginary angular momentum' therefore generates dissipation proportional to the Aharonov-Bohm phase one gains by rotating about the wavepacket center, which is given by the total magnetic flux through the wavepacket. We may therefore interpret the `imaginary angular momentum' as the tendency of the state to convert Aharonov-Bohm phases into dissipation and gain. Under this interpretation, its appearance in the effective magnetic energy becomes natural.

\emph{Gauge transformations.} 
We now discuss the effect of gauge transformations within non-Hermitian perturbation theory and its relation to our result for the effective energy \eqref{eq:Energy_mag}. Since the quasilocal Hamiltonian is a first order approximation, we begin by describing the role of gauge transformations on the general first-order perturbation theory. Suppose that $H(\lambda)$ is some parametrized Hamiltonian, where $\lambda$ represents the strength of some perturbation, such that $H(0)$ is the unperturbed Hamiltonian. Given another Hamiltonian $\tilde{H}(\lambda)$ describing the same perturbation, but in a different gauge, we may write the gauge transformation relating them as
\begin{equation}
    \tilde{H}(\lambda)=U^\dagger(\lambda)H(\lambda)U(\lambda).
\end{equation}
We note that, since the unperturbed Hamiltonian remains unchanged, we also have $U(0)=I$.

If $\ket{n^R}$ is an eigenstate of the unperturbed Hamiltonian, the first order perturbative correction to its energy is given by \cite{sternheim_PhysRevC.6.114}
\begin{equation}
    \varepsilon^{(1)}_n=\bra{n^L}\partial_\lambda H\vert_{\lambda=0}\ket{n^R}.
\end{equation}
Carrying out the calculation for the gauge-transformed Hamiltonian $\tilde{H}$ then gives
\begin{equation}
    \label{eq:gaugeEnergy}
    \tilde{\varepsilon}^{(1)}_n = \varepsilon^{(1)}_n + \bra{n^L}[H(0),\partial_\lambda U\vert_{\lambda=0}]\ket{n^R}.
\end{equation}
It is therefore immediately apparent that for the results of perturbation theory to be gauge-invariant, the difference between both calculations, given by the expectation value of the commutator, must vanish. If $\ket{n^R}$ belongs to the point spectrum of $H(0)$, such as for a localized eigenstate, this is indeed always the case. However, if $\ket{n^R}$ arises instead from the continuous spectrum, such as for extended Bloch states within a band, the above commutator may become non-trivial, making the energy correction predicted by perturbation theory gauge dependent. Mathematically, this results from the singular (operator-norm discontinuous) behavior of perturbation theory in the limit $\lambda\to 0$ \cite{nenciu_dynamics_1991}. We note that this singularity does not represent any difference between the spectra of $H$ and $\tilde{H}$, but rather an ambiguity in the association of states of the perturbed band to states of the unperturbed band. 

In the specific case of magnetic fields, the strength of the perturbation enters the Hamiltonian through the spatial rate of change of the vector potential, $\mathcal{A}(\lambda \vec{r})$. Small values of $\lambda$ then represent the limit of weak magnetic fields, where the quasi-local approximation is valid. Furthermore, we may assume that $\mathcal{A}(r_c)$ is zero, since it is always absorbed into the definition of the gauge invariant momentum $p$. A gauge transformation which preserves these properties is then of the form
\begin{equation}
    U(\lambda)=\exp\left({i\lambda\sum_i\sum_ja_{ij}(r-r_c)_i(r-r_c)_j+\mathcal{O}(\lambda^2)}\right),
\end{equation}
with $a_{ij}$ being the coefficients of an arbitrary symmetric matrix. This transformation induces a non-vanishing energy difference given by
\begin{equation}
\label{eq:gauge_error}
    \tilde{\varepsilon}^{(1)}_n - \varepsilon^{(1)}_n = \sum_{i,j}a_{ij}[(A^{LR}_j-A^{RR}_j)\partial_i\varepsilon + (A^{LR}_i-A^{RR}_i)\partial_j\varepsilon].
\end{equation}
We immediately see that if the Hamiltonian is Hermitian (normal) the energy difference vanishes, explaining why no ambiguities arise in first order Hermitian perturbation theory. In the non-Hermitian case, however, different choices of magnetic gauge would yield different results for the effective magnetic energy. Despite this, our result in Eq. \eqref{eq:Energy_mag} is correct in any gauge. This is because we have made use of the symmetric gauge, which regularizes the singularity of perturbation theory in the case of magnetic fields \cite{nenciu_asymptotic_2000}. Since the difference of the results between different gauges arises from the singular nature of perturbation theory, one will always arrive back at the result calculated in the symmetric gauge once the singularity is properly taken into account. Intuitively, the symmetric gauge holds this special property because its Landau levels admit a natural basis of eigenstates localized at the wavepacket center, which complies with the quasilocal nature of the formalism.

Finally, we remark that the effects of the singular behavior of perturbation theory are not limited to first-order calculations. Moreover, in higher orders of perturbation theory, where the effects of gauge transformations are more complex, gauge-dependent terms appear even in the Hermitian setting \cite{gao_field_2014}. This can be seen, for instance when calculating magnetic susceptibility in an arbitrary gauge \cite{gao_geometrical_2015}. This problem in higher-order Hermitian perturbation theory is also similarly solved by working in the symmetric gauge \cite{chang_berry_1996,chang_berry_2008,gao_field_2014,gao_geometrical_2015}. The non-Hermitian case is therefore unique only in that the singularity affects the results already in first-order calculations.

\emph{Conclusions.} 
In this Letter, we derived the magnetic moment of a wavepacket in a non-Hermitian periodic potential. Our results show that the magnetic moment is comprised of two contributions. The first is a coupling between the anomalous non-Hermitian Berry connection and the Lorentz force induced by the field, in analogy to similar terms which appear in the presence of an electric field. The second is a Zeeman-like contribution which is related to the self-rotation of the wave packet about its center. While this contribution is also present in the Hermitian setting, we show that its non-Hermitian generalization is no longer given by a measurable, physical angular momentum, but rather by a generalized non-Hermitian `magnetic moment' operator.

In order to relate the magnetic moment operator of the non-Hermitian setting to a physically meaningful notion of angular momentum, we give a definition and a closed form for the physical angular momentum operator which is applicable in the non-Hermitian setting. By examining our physical angular momentum operator we concluded that the Zeeman term appearing in the non-Hermitian magnetic Hamiltonian is a sum of two operators, the Hermitian physical angular momentum operator and an anti-Hermitian `imaginary angular momentum' operator. We showed that the latter of these two operators is unique to the non-Hermitian setting and expresses a non-Hermitian generalization of the Aharonov-Bohm effect in which rotation about the magnetic flux causes the wavepacket state to experience an additional norm shift (as opposed to the usual phase shifts) due to the vector potential.

Finally, we also touched upon the intricacies of applying gauge transformations to the magnetic vector potential within the framework of non-Hermitian perturbation theory. While the results of perturbation theory are known to be generally gauge dependent even in the Hermitian setting \cite{nenciu_asymptotic_2000,nenciu_dynamics_1991}, the order of perturbation at which the effect becomes relevant is different in the non-Hermitian case. Namely, it is known that in Hermitian perturbation theory gauge transformations do not affect the first order correction to the energy and instead only appear in second-order and higher corrections. In contrast, we show that the non-Hermitian theory is affected already to first order, owing to the potential disparity between left and right eigenstates in the system. This issue is regularized, as in the Hermitian theory, by working in the symmetric gauge, for which perturbation theory is particularly well behaved \cite{nenciu_asymptotic_2000}.

The theoretical formalism developed here generalizes similar methods of treating time-dependent and slowly spatially varying Hamiltonians from the Hermitian setting. The general results, Eqs. \eqref{eq:Energy_eff} and \eqref{eq:Energy_quasilocal}, presented here can therefore be adapted into additional scenarios such as lattice deformation and impurity potentials and may serve as a method for studying non-Hermitian incommensurate crystals \cite{Longhi_2019_PhysRevLett.122.237601}. Furthermore, we expect that the explicit results for the magnetic moment presented here will promote further study into the effects of non-Hermitian magnetics. In particular, we believe our work may be expanded upon to provide a semi classical theory of magnetotransport and to include the effects of spin, which was left outside of the scope of this Letter.

\begin{acknowledgments}
M.G. has been supported by the Israel Science Foundation (ISF) and the Directorate for Defense Research and Development (DDR\&D) under grant No. 3427/21, by the ISF under grant No. 1113/23, and by the US-Israel Binational Science Foundation (BSF) under Grant No. 2020072. R.I. has been supported by the U.S.-Israel Binational Science Foundation (BSF) under Grant No. 2018226. and by the Israeli Science Foundation (ISF) under Grants No.1790/18 and No. 2307/24.
R.I. is grateful for the hospitality of the Perimeter Institute, where part of this work was carried out. Research
at the Perimeter Institute is supported in part by the Government of Canada through the Department of
Innovation, Science and Economic Development and by the Province of Ontario through the Ministry of
Colleges and Universities. This work was supported by a grant from the Simons Foundation (Grant No. 1034867,
Dittrich).
\end{acknowledgments}

\nocite{balian_relation_1989}

\bibliography{NHMagneticMoment.bib}

\pagebreak
\widetext
\begin{center}
\textbf{\large Supplemental Material}
\end{center}
\setcounter{equation}{0}
\setcounter{figure}{0}
\setcounter{table}{0}

\renewcommand{\theequation}{S\arabic{equation}}
\renewcommand{\thesection}{S.\Roman{section}}

\section{Single-band approximation for time-dependent Hamiltonians}\label{app:sb_energy}

In this appendix we derive in detail the expression for the effective energy of a possibly perturbed, time-dependent Bloch Hamiltonian within the single-band approximation \eqref{eq:Energy_eff}. We begin by writing the Schr\"{o}dinger equation for a single-band state
\begin{equation}
    \label{eq:Schrodinger_SB_explicit}
    \int_{k^\prime} \dot{w}_{nk^\prime}\ket{\psi^R_{nk^\prime}(t)} 
    +\int_{k^\prime} w_{nk^\prime}\ket{\partial_t\psi^R_{nk^\prime}(t)} 
    = -i\int_{k^\prime}w_{nk^\prime}H \ket{\psi^R_{nk^\prime}(t)}.
\end{equation}
We can easily isolate the time dependence of the coefficient $w_{nk}$ at momentum $k$ by multiplying on the left with $\bra{\psi^L_{nk}}$ and attain
\begin{equation}
    \label{eq:w_dot}
    \dot{w}_{nk} = -\int_{k^\prime}w_{nk^\prime} \left( i\bra{\psi^L_{nk}(t)}H \ket{\psi^R_{nk^\prime}(t)}+\braket{\psi^L_{nk}(t)|\partial_t\psi^R_{nk^\prime}(t)}\right).
\end{equation}
Since we are interested in the effective energy of the state, we must calculate the expectation value with regards to the Hamiltonian operator. We will assume without loss of generality that the norm of the wavepacket state is $N=1$. This may be done since expectation values are always normalized by definition. Since the Hamiltonian is equivalent to the operator $i(d/dt)$, we may write the matrix element as
\begin{equation}
    \begin{split}
    \bra{W}H\ket{W} &= i\int_{kq}\bra{\psi_{nq}^R(t)}w^*_{nq} \frac{d}{dt}\left( w_{nk}\ket{\psi^R_{nk}(t)} \right)\\
    &= i\int_{kq} w^*_{nq}\dot{w}_{nk}\braket{\psi_{nq}^R(t)|\psi^R_{nk}(t)}
    + i\int_{kq} w^*_{nq}w_{nk}\braket{\psi_{nq}^R(t)|\partial_t\psi^R_{nk}(t)}.
    \end{split}
\end{equation}
It remains to solve both integrals. We begin by substituting \eqref{eq:w_dot} into the first integral to receive
\begin{equation}
    \begin{split}
     &\int_{kqk^\prime} w^*_{nq}w_{nk^\prime}\braket{\psi_{nq}^R(t)|\psi^R_{nk}(t)}\bra{\psi^L_{nk}(t)}H \ket{\psi^R_{nk^\prime}(t)} 
     -i\int_{kqk^\prime} w^*_{nq}w_{nk^\prime} \braket{\psi_{nq}^R(t)|\psi^R_{nk}(t)}\braket{\psi^L_{nk}(t)|\partial_t\psi^R_{nk^\prime}(t)} \\
     =& \int_{qk^\prime} w^*_{nq}w_{nk^\prime}\bra{\psi^R_{nq}(t)}P_nH \ket{\psi^R_{nk^\prime}(t)}
     -i\int_{k} \vert w_{nk}\vert^2 I_{nn}(k)\braket{\psi^L_{nk}(t)|\partial_t\psi^R_{nk}(t)}\\
     =& \bra{W}P_nH\ket{W} -A^{LR}_t\vert_{k=k_c},
     \end{split}
\end{equation}
where we have used the assumption that the envelope function $g(k)$ is sharply peaked around $k_c$. Additionally, we have made use of the fact that the eigenstates are Bloch functions and hence the inner product always vanishes for differing momenta, regardless of the time parameters. These properties are also used when calculating the second integral, which simply evaluates to
\begin{equation}
    i\int_{k} \vert w_{nk}\vert^2\braket{\psi_{nk}^R(t)|\partial_t\psi^R_{nk}(t)}
    = \int_{k} g(k)A^{RR}_t = A^{RR}_t\vert_{k=k_c}.
\end{equation}
The sum of both integrals is precisely the result quoted in the main text.

\section{Effective energy of quasi-local Hamiltonian}\label{app:quasilocal_energy}
In this appendix we derive the effective energy of a general Hamiltonian in the quasi-local approximation \eqref{eq:Energy_quasilocal}. That is, we calculate the effective energy \eqref{eq:Energy_eff} in the particular case of the quasi-local Hamiltonian \eqref{eq:quasilocal_Ham}
\begin{equation}
    \mathcal{E} = \langle H(r_c) \rangle_W
    - (A^{LR}_t-A^{RR}_t) 
    + \left\langle \frac{1}{2} P_n \left[(\hat{r}-r_{c})\frac{\partial H}{\partial r_{c}}+\frac{\partial H}{\partial r_{c}}(\hat{r}-r_{c})\right] \right\rangle_W.
\end{equation}
The first term is plainly the unperturbed band energy. The second term expresses a contribution due to a time dependence of the band Hamiltonian. In the quasi-local approximation, this effect is a consequence of the dependence of the Hamiltonian on the wavepacket center $r_c$, which itself evolves in time. We may therefore immediately write
\begin{equation}
    \label{eq:velocity_energy}
    (A^{LR}_t-A^{RR}_t)=\dot{r}_c \cdot (A^{LR}_{r_c}-A^{RR}_{r_c}).
\end{equation}
It remains to calculate the expectation value in the third and final term. For the sake of simplifying notation, we perform the calculation for a one-dimensional system. The generalization to any spatial dimension may be trivially obtained by repeating the calculation for each coordinate separately.

We begin the calculation by writing an expression for the (bi-orthogonal) matrix elements of the operators involves. For the derivative of the Hamiltonian we have
\begin{equation}
    \label{eq:dH_matrix_element}
    \begin{split}
    \bra{\psi^L_{n^\prime k^\prime}(r_c)}\frac{\partial H}{\partial r_{c}}\ket{\psi^R_{nk}(r_c)} =
    \delta(k-k^\prime)\delta_{nn^\prime}\partial_{r_c}\varepsilon_{nk} 
    + \delta(k-k^\prime)(\varepsilon_{nk} - \varepsilon_{n^\prime k})\braket{u^L_{n^\prime k}|\partial_{r_c}u^R_{nk}},
    \end{split}
\end{equation}
Where we have again used the orthogonality of modes with different momenta and the fact that, by duality of the eigenbases, derivatives of the form $\partial_{r_c}\braket{u^L_{n^\prime k^\prime}|u^R_{nk}}$ always vanish.
For the position operator, the form of a generic matrix element has been calculated before \cite{balian_relation_1989,silberstein_berry_2020}. Here we are interested specifically in the bi-orthogonal matrix element, which is given by
\begin{equation}
    \label{eq:r_matrix_element}
    \bra{\psi^{L}_{n^\prime k^\prime}}\hat{r}\ket{\psi^R_{nk}}= i\delta(k-k^\prime)\braket{u^{L}_{n^\prime k} | \partial_k u^R_{nk}}
    -i\partial_k\delta(k-k^\prime)\delta_{n^\prime n}.
\end{equation}

We now set out to calculate the expectation value itself. Writing it explicitly and making use of the above matrix elements we may split the expectation value into three separate contributions
\begin{equation}
    \begin{split}
    &\frac{1}{2}\int_{kk^\prime q} w^*_{nk^\prime}w_{nk} \braket{\psi^R_{nk^\prime}|\psi^R_{nq}}
    \left( \bra{\psi^L_{nq}}  (\hat{r}-r_{c})\frac{\partial H}{\partial r_{c}} \ket{\psi^R_{nk}}
    +\bra{\psi^L_{nq}}  \frac{\partial H}{\partial r_{c}}(\hat{r}-r_{c}) \ket{\psi^R_{nk}} \right) \\
    =& \frac{1}{2}\int_{kk^\prime} w^*_{nk^\prime}w_{nk} I_{nn}(k^\prime)
    \left( \bra{\psi^L_{nk^\prime}}  (\hat{r}-r_{c})\frac{\partial H}{\partial r_{c}} \ket{\psi^R_{nk}}
    +\bra{\psi^L_{nk^\prime}}  \frac{\partial H}{\partial r_{c}}(\hat{r}-r_{c}) \ket{\psi^R_{nk}} \right)\\
    =& \frac{1}{2}\sum_m\int_{kk^\prime q} w^*_{nk^\prime}w_{nk} I_{nn}(k^\prime)
    \left( \bra{\psi^L_{nk^\prime}}  (\hat{r}-r_{c})\ket{\psi^R_{mq}}\bra{\psi^L_{mq}}\frac{\partial H}{\partial r_{c}} \ket{\psi^R_{nk}}
    +\bra{\psi^L_{nk^\prime}}  \frac{\partial H}{\partial r_{c}}\ket{\psi^R_{mq}}\bra{\psi^L_{mq}}(\hat{r}-r_{c}) \ket{\psi^R_{nk}} \right) \\
    =& \frac{1}{2}\sum_m\int_{kk^\prime q} w^*_{nk^\prime}w_{nk} I_{nn}(k^\prime)
     \bra{\psi^L_{nk^\prime}}  (\hat{r}-r_{c})\ket{\psi^R_{mq}} \delta(k-q)(\varepsilon_{nk} - \varepsilon_{mk})\braket{u^L_{m k}|\partial_{r_c}u^R_{nk}} \\
    & +\frac{1}{2}\sum_m\int_{kk^\prime q} w^*_{nk^\prime}w_{nk} I_{nn}(k^\prime) \delta(q-k^\prime)(\varepsilon_{mk^\prime} - \varepsilon_{nk^\prime})\braket{u^L_{nk^\prime}|\partial_{r_c}u^R_{mk^\prime}}\bra{\psi^L_{mq}}(\hat{r}-r_{c}) \ket{\psi^R_{nk}} \\
    & +\frac{1}{2}\sum_m\int_{kk^\prime q} w^*_{nk^\prime}w_{nk} I_{nn}(k^\prime)
    \left( \bra{\psi^L_{nk^\prime}}  (\hat{r}-r_{c})\ket{\psi^R_{mq}}\delta(k-q)\delta_{nm}\partial_{r_c}\varepsilon_{nk}
    +\delta(q-k^\prime)\delta_{nm}\partial_{r_c}\varepsilon_{nk^\prime}\bra{\psi^L_{mq}}(\hat{r}-r_{c}) \ket{\psi^R_{nk}} \right)\\
    =& \frac{1}{2}\sum_m\int_{kk^\prime} w^*_{nk^\prime}w_{nk} I_{nn}(k^\prime)
     \bra{\psi^L_{nk^\prime}}  (\hat{r}-r_{c})\ket{\psi^R_{mk}} (\varepsilon_{nk} - \varepsilon_{mk})\braket{u^L_{m k}|\partial_{r_c}u^R_{nk}} \\
    & +\frac{1}{2}\sum_m\int_{kk^\prime} w^*_{nk^\prime}w_{nk} I_{nn}(k^\prime) (\varepsilon_{mk^\prime} - \varepsilon_{nk^\prime})\braket{u^L_{nk^\prime}|\partial_{r_c}u^R_{mk^\prime}}\bra{\psi^L_{mk^\prime}}(\hat{r}-r_{c}) \ket{\psi^R_{nk}} \\
    & +\frac{1}{2}\int_{kk^\prime} w^*_{nk^\prime}w_{nk} I_{nn}(k^\prime)
    \left( \bra{\psi^L_{nk^\prime}}  (\hat{r}-r_{c})\ket{\psi^R_{nk}}\partial_{r_c}\varepsilon_{nk}
    +\partial_{r_c}\varepsilon_{nk^\prime}\bra{\psi^L_{nk^\prime}}(\hat{r}-r_{c}) \ket{\psi^R_{nk}} \right), \\
    \end{split}
\end{equation}
where we have used the resolution of identity $1=\sum_m \int_q \ket{\psi^R_{mq}}\bra{\psi^L_{mq}}$. We denote the above three integrals as $I_1, I_2, I_3$ respectively, and evaluate each in turn. Beginning with the first integral, we observe that it vanishes whenever $m=n$, and hence only inter-band terms in the position operator may contribute. We may therefore express the first integral as
\begin{equation}
    \begin{split}
    I_1 &= \frac{1}{2}\sum_{m}\int_{kk^\prime} w^*_{nk^\prime}w_{nk} I_{nn}(k^\prime)
     i\delta(k-k^\prime)\braket{u^{L}_{n k^\prime} | \partial_k u^R_{mk}} (\varepsilon_{nk} - \varepsilon_{mk})\braket{u^L_{m k}|\partial_{r_c}u^R_{nk}}\\
     &= -\frac{1}{2}\sum_{m}\int_{k} \vert w_{nk}\vert^2 I_{nn}(k)
     i\bra{\partial_k u^{L}_{n k}}(\varepsilon_{nk} - \varepsilon_{mk})\ket{u^R_{mk}} \braket{u^L_{m k}|\partial_{r_c}u^R_{nk}} \\
     &= -\frac{i}{2} \bra{\partial_k u^{L}_{n k}} \varepsilon_{nk} - H(k) \ket{\partial_{r_c}u^R_{nk}}\vert_{k=k_c}.
     \end{split}
\end{equation}
By a nearly identical reasoning, $I_2$ may also be evaluated as
\begin{equation}
    I_2 = \frac{i}{2} \bra{\partial_{r_c} u^{L}_{n k}} \varepsilon_{nk} - H(k) \ket{\partial_k u^R_{nk}}\vert_{k=k_c}.
\end{equation}
It remains to calculate $I_3$. While a direct calculation using \eqref{eq:r_matrix_element} is perfectly possible, it is much less cumbersome to instead make the observation that the integral may be expressed as a sum of two expectation values
\begin{equation}
    I_3 = \frac{1}{2}\left( \langle P_n (\hat{r} - r_c)\partial_{r_c}\varepsilon_n(\hat{k})\rangle_W
    + \langle P_n \partial_{r_c}\varepsilon_n(\hat{k})(\hat{r} - r_c)\rangle_W\right).
\end{equation}
We note that since the expectation values are taken with respect to a state confined to the $n$th band, it is indeed possible to treat the (derivative of) the band energy as a function of the momentum operator. Recalling that the projection $P_n$ acts trivially on $\ket{W}$, and commutes with $\hat{k}$, and that by definition $r_c = \langle \hat{r}\rangle_W$, we may write the following useful identity
\begin{equation}
    \langle P_n (\hat{r} - r_c)f(\hat{k})\rangle_W
    + \langle P_n f(\hat{k})(\hat{r} - r_c)\rangle_W
    = 2\left( \langle f(\hat{k}) P_n \hat{r}\rangle_W
    -\langle f(\hat{k})\rangle_W \langle P_n \hat{r}\rangle_W +\frac{i}{2} \langle f^\prime(\hat{k})\rangle_W \right) -2 \langle f(\hat{k})\rangle_W \langle Q_n \hat{r}\rangle_W,
\end{equation}
where $Q_n = 1-P_n$ is the complement of the band projection. The expectation values on the right hand side were all calculated in general in a previous work \cite{alon_quantum_2024}, from which we may immediately conclude that the term in parenthesis vanishes identically and that $I_3$ is therefore given by
\begin{equation}
    I_3 = \partial_{r_c}\varepsilon_{nk}(A^{LR}_k-A^{RR}_k)\vert_{k=k_c}.
\end{equation}
Combining $I_1$, $I_2$, and $I_3$ together with \eqref{eq:velocity_energy} yields the result quoted in the main text.

\end{document}